\documentclass[%
 aip,
 amsmath,amssymb,
 reprint,%
]{revtex4-1}

\usepackage{graphicx}
\usepackage{dcolumn}
\usepackage{bm}

\usepackage[utf8]{inputenc}
\usepackage[T1]{fontenc}
\usepackage{mathptmx}
\usepackage{etoolbox}
\usepackage[dvips]{color}
\usepackage{hyperref}

\makeatletter
\def\@email#1#2{%
 \endgroup
 \patchcmd{\titleblock@produce}
  {\frontmatter@RRAPformat}
  {\frontmatter@RRAPformat{\produce@RRAP{#1{#2}}}\frontmatter@RRAPformat}
  {}{}
}%

\let\@fnsymbol\@fnsymbol@latex
\@booleanfalse\altaffilletter@sw

\makeatother

\begin{document}

\preprint{AIP/123-QED}

\title{Second-harmonic generation in germanium-on-insulator from visible to telecom wavelengths}
\affiliation{School of Electrical and Electronic Engineering, Nanyang Technological University,\\ 50 Nanyang Avenue, Singapore 639798, Singapore\\}%
\author{Yadong Wang$^{*}$}
\author{Daniel Burt$^{*}$}
\author{Kunze Lu}
\author{Donguk Nam$^{\dagger}$}
\email[{$^{\dagger}$} Corresponding author: ]{dnam@ntu.edu.sg}
\email[{$^{*}$} ]{YW and DB equally contributed to this work}
\date{\today}
\begin{abstract}

The second-order $\chi^{2}$ process underpins many important nonlinear optical applications in the field of classical and quantum optics. Generally, the $\chi^{2}$ process manifests itself only in a non-centrosymmetric dielectric medium via an anharmonic electron oscillation when driven by an intense optical field. Due to inversion symmetry, group-IV semiconductors like silicon (Si) and germanium (Ge) are traditionally not considered as ideal candidates for second-order nonlinear optics applications. Here, we report the experimental observation of the second-harmonic generation (SHG) in a Ge-on-insulator (GOI) sample under femtosecond optical pumping. Specially, we report the first-time measurement of the SHG signal from a GOI sample in the telecom S-band by pumping at $\sim$$3000$~nm.

\end{abstract}

\maketitle
\section{\label{sec:level1}Introduction}
The first observation of second-harmonic generation (SHG) by Franken \textit{et al.} in 1961 \cite{Franken} has always been considered as the foundation for the field of nonlinear optics. Nowadays, SHG has been widely applied in many important scientific applications such as ultra-short pulse measurements \cite{Kane, Nomura} and optical microscopy in biology \cite{Campagnola,Chen1}. More recently, the attention on SHG has been significantly increasing in the realm of modern quantum optics using entangled photon pairs \cite{Brendel,Dousse,LiuJ}, which can be generated by the reversed-SHG process, known as spontaneous parametric down-conversion (SPDC) \cite{Arnaut, Walther, Guo}.

Unfortunately, such second-order nonlinear processes enabled by a finite $\chi^{2}$ susceptibility are normally prohibited in foundry compatible semiconductor materials like silicon (Si) and germanium (Ge) due to their centrosymmetric unit cell. Numerous schemes have been proposed to enhance the second-order nonlinearity in Si photonics. Examples include strain engineering of Si \cite{Jacobsen, Cazzanelli}, resonant excitation of crystalline Si nanoparticles \cite{Makarov}, and application of an electric field to Si waveguides \cite{Timurdogan}. In contrast, for Ge, there is only a theoretical prediction of the mid-infrared (MIR) SHG in a strained Ge waveguide \cite{Leonardis}. Since Ge is already on the list of materials in the Si photonics foundry and has recently gained prominence for on-chip lasers \cite{Rong, Petykiewicz, Bao, Qi, Joo, Kim} and MIR waveguides \cite{Soref,Kang}. It is desirable to enable and enhance second-order nonlinear optical processes in Ge to widen the functionality of this important material.
 
In this Article, we demonstrate the experimental observation of SHG in visible and telecom S-band wavelengths from a germanium-on-insulator (GOI) sample. A tunable femtosecond pulsed laser was used to pump the sample from near-infrared (NIR) to MIR spectral ranges up to $3000$ nm. Typical quadratic power dependence of the SHG signal was confirmed both in visible ($\sim$$550$ nm) and S-band ($\sim$$1500$ nm) wavelengths. By systematically tuning the pump wavelength, we also observe that the SHG wavelength is always double the photon energies of the pump photons, further confirming the validity of our SHG signals. We believe that the underlying physics of this SHG in Ge could be explained by the surface effect induced by symmetry breaking at the material interface \cite{Hollering,Heinz, Chen}. This result shows the potential to produce entangled photon pairs using SPDC in Ge that holds the promise to the realization of monolithic quantum photonic-integrated circuits (PICs). More importantly, our first experimental report on generating telecom photons via a nonlinear optical process will pave the way for NIR and MIR nonlinear optics applications in Si photonics.

\begin{figure}[b]	
	\includegraphics[width=\linewidth]{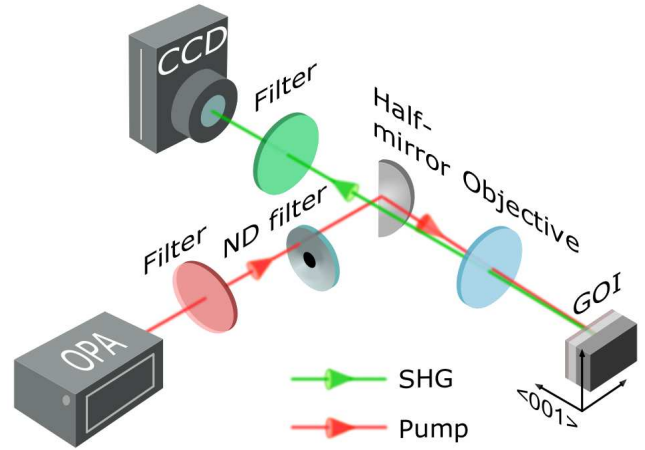}
	\caption{Schematics of the free-space SHG experimental setup, showing the excitation pump optical path (red beam) from the OPA to the GOI sample, and the SHG signal path (green beam) from the sample to the CCD. A $40\times$ reflective objective with a $0.5$ NA is used to focus the excitation pump from NIR to MIR wavelengths onto the GOI sample.}
	\label{fig:Fig1}
\end{figure}

\section{\label{sec:level2}NONLINEAR OPTICAL CHARACTERIZATION}

Our SHG experiments were performed in a home-built microscope spectroscopy setup, which is schematically depicted in Fig.\ref{fig:Fig1}. Here, we used a femtosecond pulse optical parametric amplifier (OPA) with a repetition rate of 100 kHz as a tunable excitation pump. The OPA generated a tunable femtosecond laser pulses with wavelengths between $532$ to $3000$ nm. Using the collinear nature of the OPA output, we carefully aligned and optimized the $1000$ to $3000$ nm pumps (red beam in Fig.\ref{fig:Fig1}) by using its visible beam at $532$ nm as the reference optical path. To spectrally block the unwanted OPA idlers lower than the primary pump, optical filters were placed in front of the OPA output. A variable neutral density (ND) filter was used to attenuate the intense pump in the path. Then the attenuated optical beam was directed to an objective lens via a silver-coated half-mirror. We selected a reflective objective with a $40$$\times$ magnification and $0.5$ numerical aperture (NA) to focus the pump on the GOI sample. It allowed for both the optimal transmission of the pump from NIR to MIR and the collection of the SHG signal ranging from visible to telecom wavelengths. To measure the SHG spectra with a spectral resolution of $0.28$ nm, we used two separate $1$D array detectors for visible and NIR spectrum measurement integrated on a spectrograph.

For this experimental study, we prepared a GOI sample, in which a Ge wafer is bonded to an oxidized Si substrate. The GOI was created using a direct wafer bonding technique \cite{Nam,Nam2,Burt}. The thickness of the GOI and the buried oxide layers were determined to be $350$ nm and $1000$ nm, respectively, by using ellipsometry (not shown here).

\section{\label{sec:level3}SECOND-HARMONIC GENERATION IN THE VISIBLE WAVELENGTHS}

\begin{figure}[t]
	\includegraphics[width=\linewidth]{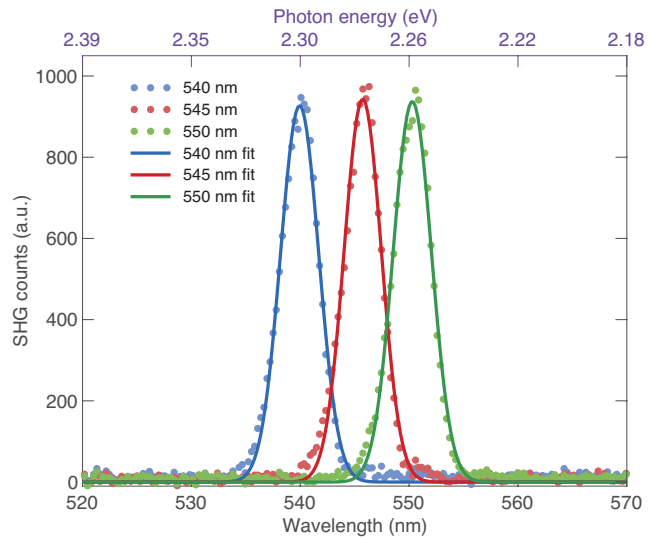}
	\caption{The SHG spectra measured on the GOI sample induced by the OPA pumping at the wavelengths of $1080$, $1090$, and $1100$ nm. With the pump power fixed at $2.12$ mW, the measured SHG spectra show their peak positions at $540$, $545$, and $550$ nm as drawn in blue, red, and green dots, respectively. The coloured lines are the corresponding Gaussian fits.}
	\label{fig:Fig2}
\end{figure}

To demonstrate the $\chi^{(2)}$ mediated SHG in Ge samples, we first performed the real-time spectral measurement on the GOI using the setup shown in Fig.\ref{fig:Fig1}. The OPA pump wavelength was initially set at $1100$ nm, and its power onto the sample was adjusted to $2.12$ mW. Here, the photon energy of the excitation beam ($\sim$$1.13$ eV) is higher than the direct and indirect bandgap energies of Ge similar to the choice of the excitation pump used by Hollering \textit{et al} \cite{Hollering}. With the exposure time set to $1$ s for the visible CCD detector, a distinct SHG peak (green dots) centered at $550$ nm can be clearly observed on the spectrum in Fig.\ref{fig:Fig2}. Next, we switched the OPA wavelength to $1090$ and $1080$ nm while maintaining the same power as the previous measurement. The SHG signals shifted correspondingly to $545$ (red dots) and $540$ nm (blue dots). The Gaussian fittings of the measured SHG were also plotted. Note that we carefully chose the measurement window to be centered at $550$ nm with a spectral width of $\sim$$170$ nm for these measurements. A $900$-nm short-pass filter was placed in front of the CCD to block the intense pump laser and its idler. To confirm that this signal at around $550$ nm is the SHG response in Ge, we carefully repeated the same test on a bare Si substrate, which did not show any signal with the same measurement parameters. It is worth noting that the same SHG response was observed in both bulk Ge and epitaxially grown Ge on Si, proving that the SHG is not exclusive to the GOI sample.

\begin{figure}[b]
	\includegraphics[width=\linewidth]{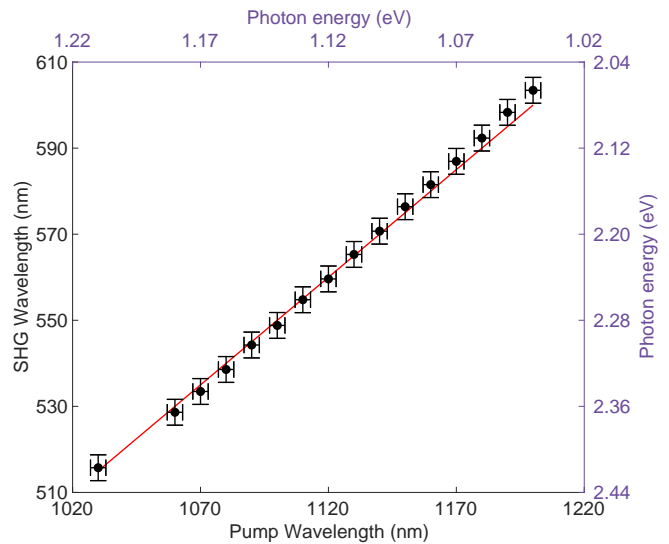}
	\caption{The measured SHG wavelength as a function of the pump wavelength by tuning the pump wavelengths from $1030$ to $1200$ nm while keeping the average pump power to $2.12$ mW. The black dots are the measurement data with a $\pm5$ nm error and the red line indicates the expected SHG relationship calculated with the experimental parameters.}
	\label{fig:Fig3}
\end{figure}

Figure \ref{fig:Fig3} illustrates the measured SHG wavelength as a function of the pump wavelength. For this experiment, we linearly swept the OPA wavelength from $1030$ to $1200$ nm at a fixed power of $2.12$ mW, and captured the corresponding SHG spectra with a $1$ s exposure time. Then, we numerically extracted the SHG peak position by fitting the signal with a Gaussian profile. The top x-axis and right y-axis in purple in Fig.\ref{fig:Fig3} are the equivalent axes in photon energy. The error bars of $\pm5$ nm were evaluated based on the spectrograph gratings and specifications of the OPA. There is a good agreement between the extracted data and the theoretical line in red, confirming the validity of the SHG.

\begin{figure}[b]
	\includegraphics[width=\linewidth]{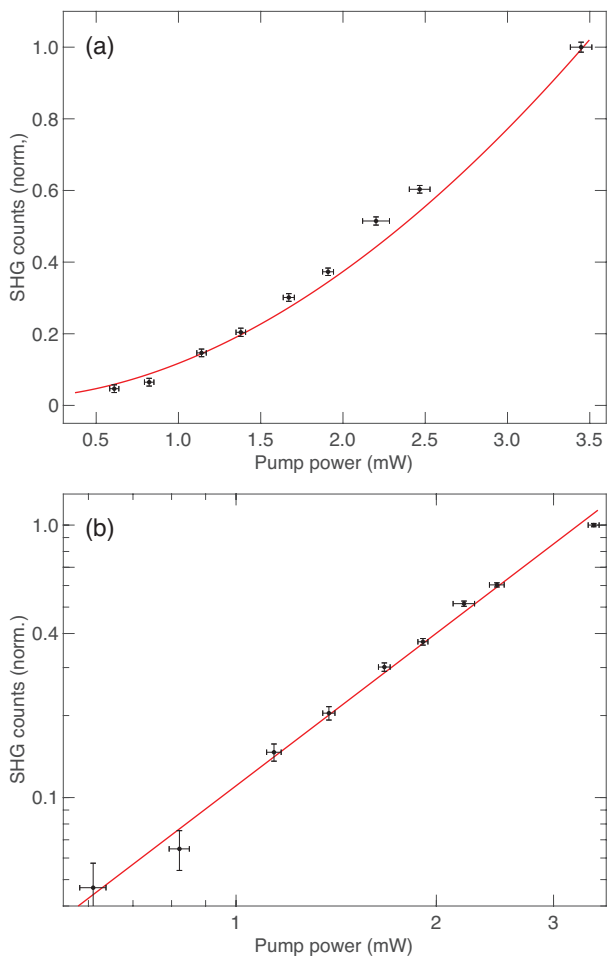}
	\caption{(a) Linear plot of the normalized SHG counts as a function of pump powers between $0.61$ to $3.45$ mW at the excitation wavelength of $1100$ nm ($1.13$ eV). The black dots show the measured data, and the red curve is the second-order polynomial fitting.  (b) Double logarithmic presentation of the quadratic power dependence with an estimated exponent of $1.86\pm0.32$. }
	\label{fig:Fig4}
\end{figure}

In Fig.\ref{fig:Fig4}(a), we demonstrated the well-known signature of SHG in which the signal intensity follows a quadratic relationship with the excitation pump power \cite{Boyd}. Here, by fixing the pump wavelength at $1100$ nm, we linearly increased the pump power from $0.61$ to $3.45$ mW. The measured SHG signal (black dots) in the normalized counts exhibited a clear quadratic power dependence in Fig.\ref{fig:Fig4}(a). The uncertainties (black error bars) in the pump power were determined by the standard deviation (SD) of a $100$ times repeated measurements at each pump power, and the errors in the SHG counts were estimated based on the calculated SD of the background signal noise with the same experimental settings. A second-order polynomial fitting with a $95\%$ confidence interval is displayed as the red curve. We also show the power law as a double logarithmic plot in Fig.\ref{fig:Fig4}(b) \cite{Chen,Timurdogan}. These results demonstrate that the SHG measured in the GOI is in excellent agreement with the theory, with an estimated exponent of $1.86$ $\pm$ $0.32$ (red line).

\section{\label{sec:level5}SECOND-HARMONIC GENERATION IN THE TELECOM S-BAND}

\begin{figure}[pb]
	\includegraphics[width=\linewidth]{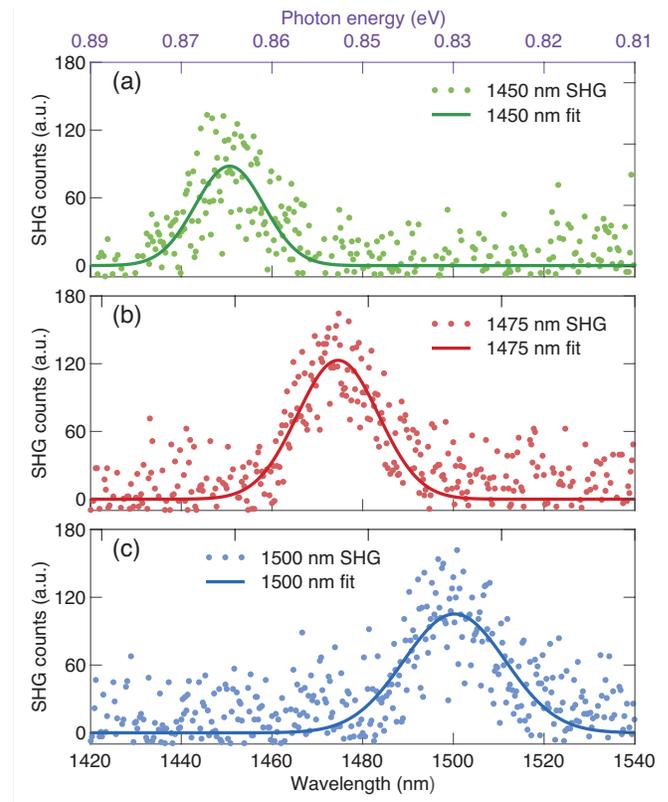}
	\caption{The SHG spectra centered  at (a) $1450$, (b) $1475$ , and (c) $1500$ nm measured on the GOI with the OPA pump at $2900$, $2950$, and $3000$ nm, respectively. The coloured dots are the measured spectra, whereas the solid curves show their corresponding Gaussian fits.}
	\label{fig:Fig5}
\end{figure}

The GOI at room temperature typically displays a high intrinsic material absorptions between the visible to NIR with a cut-off wavelength around $1500$ nm \cite{Stillman, Gosciniak}. For the SHG signal in the visible range enabled by the NIR pump (see Figs.\ref{fig:Fig2}-\ref{fig:Fig4}), the effect of the material absorption is inevitable. Hence, we decided to move the excitation pump into the MIR at a wavelength of $3000$ nm, where the material absorption can be negligible for both pump and SHG. Since the pump photons have a much smaller energy than the bandgap energy of Ge, the electrons are only excited into virtual levels within the bandgap. 
	
The optical setup for the MIR pumping is nearly the same as for the NIR pumping described in Fig.\ref{fig:Fig1}. A $2400$-nm long-pass filter was placed at the OPA output to only transmit the desired pump photons at $\sim$$3000$ nm.  First, with a $2900$-nm pump at $4.5$ mW and a CCD exposure time set at $60$ s, the SHG spectrum at $1450$ nm was measured as the green dots shown in Fig.\ref{fig:Fig5}(a). Subsequently, we increased the pump wavelength in a $50$-nm increment, and plotted the corresponding SHG spectra at $1475$ and $1500$ nm in Figs.\ref{fig:Fig5}(b) and (c), respectively. The SHG spectra at the telecom wavelength are spectrally noisier than that in the visible wavelength shown in Fig.\ref{fig:Fig2}. This is accredited to the lower detection efficiency and higher dark currents of the InGaAs detector compared to the Si detector. We carefully confirmed the reproducibility of the SHG signals at the telecom S-band by repeating the same measurement at different areas on the GOI sample. To the best of our knowledge, this is the first experimental demonstration of the SHG signal in the technologically important telecom wavelength in Ge.
\linespread{-0.5}
\vspace{-5mm}
\section{\label{sec:level6}Conclusion}

To conclude, we reported the first experimental observation of the tunable SHG in a GOI sample ranging from visible to telecom wavelengths. We confirmed the validity of the SHG signals by studying the SHG wavelengths tuned by the pump wavelengths and also by presenting the characteristic quadratic power dependence. Whilst our current experimental setup has enabled the proof-of-principle SHG in Ge, there are a number of improvements that are left for future work. For example, we could improve the current setup to enable the SHG mapping on the Ge sample which potentially could provide more insights. We believe that our work paves the way toward developing nonlinear Si photonic devices operating in the telecom wavelengths.
\vspace{-5mm}
\begin{acknowledgments}
The research of the project was in part supported by Ministry of Education, Singapore, under grant AcRF TIER $1$ (RG $115$/$21$). The research of the project was also supported by Ministry of Education, Singapore, under grant AcRF TIER $2$ (MOE$2018$-T$2$-$2$-$011$ (S)). This work is also supported by National Research Foundation of Singapore through the Competitive Research Program (NRF-CRP$19$-$2017$-$01$). This work is also supported by National Research Foundation of Singapore through the NRF-ANR Joint Grant (NRF$2018$-NRF-ANR$009$ TIGER). This work is also supported by the iGrant of Singapore A*STAR AME IRG (A$2083$c$0053$). The authors would like to acknowledge and thank the Nanyang NanoFabrication Centre (N$2$FC).
\end{acknowledgments}

\section*{Data Availability Statement}

The data that support the findings of this study are available from the corresponding author upon reasonable request.

\newcommand{\enquote}[1]{``#1''}


\end{document}